\documentclass[12pt]{article}

\newcommand{\beq}{\begin{equation}}
\newcommand{\eeq}{\end{equation}}
\newcommand{\beqa}{\begin{eqnarray}}
\newcommand{\eeqa}{\end{eqnarray}}
\newcommand{\beqar}{\begin{eqnarray*}}
\newcommand{\eeqar}{\end{eqnarray*}}

\newcommand{\al}{\alpha}
\newcommand{\be}{\beta}

\def\non          {\nonumber}

\def\Tr           {\mbox{\rm Tr}\,}
\def\STr          {\mbox{\rm STr}\,}
\def\Str          {\mbox{\rm Str}\,}

\def\cd           {{\cdot}}
\def\ran          {\rangle}
\def\lan          {\langle}
\def\fsH	{H\!\!\!\!/\,}

\newcommand{\eps}{\epsilon}

\newcommand{\inn}{\!\cdot\!}

\newcommand{\lam}{\lambda}

\newcommand{\z}{\zeta}

\newcommand{\eg}{{\it e.g.,}\ }
\newcommand{\ie}{{\it i.e.,}\ }
\newcommand{\labell}[1]{\label{#1}} 
\newcommand{\reef}[1]{(\ref{#1})}
\newcommand\prt{\partial}

\newcommand\bz{\bar{z}}

\begin{document}
\baselineskip 18pt%
\begin{titlepage}
\vspace*{1mm}%
\hfill
\vbox{
    \halign{#\hfil         \cr
          IPM/P-2008/067  \cr
           } 
      }  
\vspace*{20mm}
\vspace*{15mm}%

\center{ {\bf \Large More on WZ action of  non-BPS branes 
}}\vspace*{3mm} \centerline{{\Large {\bf  }}}
\vspace*{5mm}
\begin{center}
{ Mohammad R. Garousi and Ehsan Hatefi}

\vspace*{0.8cm}{ {Department of Physics, Ferdowsi University of Mashhad,\\
P.O. Box 1436, Mashhad, Iran}}\\

 {  School of Physics, 
 Institute for research in fundamental sciences (IPM), \\
 P.O.Box 19395-5531, Tehran, Iran}

E-mails : {{ garousi@mail.ipm.ir, hatefi@ipm.ir}}\\
\vspace*{1.5cm}
\end{center}
\begin{center}{\bf Abstract}\end{center}
\begin{quote}

We calculate the disk level S-matrix element of one Ramond-Ramond, two gauge field and one tachyon vertex operators in the world volume of non-BPS branes. We then find the  momentum expansion of this amplitude and show that the infinite tachyon/massless poles and the contact terms of this amplitude can be reproduced by  the tachyon DBI  and the WZ actions, and by their higher derivative corrections. 

\end{quote}
\end{titlepage}

\section{Introduction}

Spectrum of non-BPS branes has tachyon, massless and infinite tower of massive states. Even though the mass scale of tachyon and the massive states are the same, there are various arguments that indicate  there must be an effective theory for non-BPS branes that includes only the tachyon and the massless states \cite{Sen:2004nf}. The effective theory  should have two parts, \ie 
\beqa
S_{non-BPS}&=&S_{DBI}+S_{WZ}\labell{nonbps}\nonumber\eeqa
where $S_{DBI}/S_{WZ}$ should be an extension of DBI/WZ action of BPS branes in which the tachyon mode of non-BPS brane
are included appropriately. 

One method for finding these effective actions is the BSFT. In this formalism the kinetic term of tachyon appears in the DBI part  as \cite{Kraus:2000nj,Takayanagi:2000rz}
\beqa
S_{DBI}&\sim&\int d^{p+1}\sigma\, e^{-2\pi T^2}\left(F(2\pi\alpha'D^aTD_aT)+\cdots\right),\qquad\qquad F(x)=\frac{4^xx\Gamma(x)^2}{2\Gamma(2x)}
\eeqa
where dots refer to the gauge field couplings. When tachyon is zero, they are given by the DBI action. The WZ term in this formalism is given by \cite{Kraus:2000nj,Takayanagi:2000rz}
\beqa
S_{WZ}&=&\mu_p' \int_{\Sigma_{(p+1)}} C \wedge \Str e^{i2\pi\alpha'\cal F}\labell{WZ'}\eeqa 
 in which the curvature of superconnection is 
 \begin{displaymath}
i{\cal F} = \left(
\begin{array}{cc}
iF -\beta'^2 T^2 & \beta' DT \\
\beta' DT & iF -\beta'^2T^2 
\end{array}
\right) \ ,
\non\end{displaymath}
 and  $\beta'$ is a  normalization constant with dimension $1/\sqrt{\alpha'}$. 

Another method for studying the effective action of non-BPS branes is the S-matrix method. In this formalism the kinetic term of tachyon appear in the DBI action as \cite{Garousi:2008tn,Garousi:2008ze}
\beqa
S_{DBI}&\sim&\int
d^{p+1}\sigma \STr\left(\frac{}{}V({ T^iT^i})\sqrt{1+\frac{1}{2}[T^i,T^j][T^j,T^i])}\right.\labell{nonab} \\
&&\qquad\qquad\left.
\times\sqrt{-\det(\eta_{ab}
+2\pi\alpha'F_{ab}+2\pi\alpha'D_a{ T^i}(Q^{-1})^{ij}D_b{ T^j})} \right)\,,\nonumber\eeqa  where $V({T^iT^i})=e^{-\pi{ T^iT^i}/2}$, and 
\beqa
Q^{ij}&=&I\delta^{ij}-i[T^i,T^j]\eeqa
The superscripts $i,j=1,2$, \ie $T^1=T\sigma_1$, $T^2=T\sigma_2$ and there is no sum over $i,j$. $\sigma_1$ and $\sigma_2$ are the Pauli matrices. After expanding the square roots one should choose two of the tachyons to be $T^2$ and the others to be $T^1$.  The trace in above equation
must be completely symmetric between all  matrices
of the form $F_{ab},D_a{ T^i}$, $[T^i,T^j]$ and individual
${ T^i}$ of the  potential $V(T^iT^i)$. The above action is consistent with the momentum expansion of the S-matrix element of four tachyons,  the S-matrix element of one RR and three tachyons \cite{Garousi:2008tn,Garousi:2008ze} and with the momentum expansion of the S-matrix element of four tachyons and one gauge field \cite{Garousi:2008nj}. Around the stable point of the tachyon potential, the above action reduces to the usual tachyon DBI action \cite{Sen:1999md,Garousi:2000tr,Bergshoeff:2000dq,Kluson:2000iy} with potential $T^4V(T^2)$.  The WZ part in this formalism, on the other hand, is given by the same WZ action as in the BSFT in which the normalization of tachyon is $\beta'=\frac{1}{\pi}\sqrt{\frac{6\ln(2)}{\alpha'}}$  for non-BPS branes \cite{Garousi:2008tn} and $\beta=\frac{1}{\pi}\sqrt{\frac{2\ln(2)}{\alpha'}}$  for  brane-anti-brane \cite{Garousi:2007fk}.  In fact the structure of superconnection in the WZ part has been  found first by the S-matrix method in \cite{Kennedy:1999nn}.

In this paper, we would like to examine the above actions  with  the S-matrix element of one RR, two gauge fields and one tachyon. So in the next section we calculate this S-matrix element in string theory. In section 3, we find the momentum expansion of the amplitude and compare it  with the various Feynman amplitudes resulting from  the couplings in the above actions and their higher derivative extensions.

\section{The four-point amplitude}
To calculate a S-matrix element, one needs to choose the picture of the vertex operators  appropriately. The sum of the superghost charge must be -2 for disk level amplitude.  On the other hand, the vertex operators of a  non-BPS D-brane carry internal CP matrix \cite{Sen:1999mg}. Using the fact that the picture changing operator also carries internal CP matrix $\sigma_3$ \cite{DeSmet:2000je}, one realizes  that a vertex operator carries different internal CP matrices depending on its superghost charges. It has been speculated in \cite{Garousi:2008ze} that the  S-matrix element of $n$ external states is independent of the choice of the picture of the external states only  when one includes the internal CP matrix in the vertex operators. 

When tachyon is set to zero, the effective field theory of non-BPS branes/brane-antibrane must reduce to  the effective field theory of BPS branes in which there is no internal CP matrix. This indicates that  the massless fields in the  effective field theory of non-BPS branes/brane-antibrane must carry identity internal CP matrix. For example, the RR field in the effective field theory of brane-antibrane must carry identity matrix because when tachyon is set to zero the WZ action of brane-antibrane  reduces to the WZ action of two stable BPS branes. This internal CP matrix is contributed to the  RR vertex operator in $(0)$-picture. In other pictures, the internal CP matrix is different depending on its superghost picture. It has been shown in \cite{Sen:1999mg} that the RR vertex operator of  a non-BPS brane in $(0)$-picture must carry the internal CP matrix $\sigma_1$. This vertex operator in $(-1)$-picture then must carry the internal matrix $\sigma_3\sigma_1$.

Hence, the S-matrix element of one RR field, two gauge fields and one tachyon
in the world volume of non-BPS branes  is given by the following correlation function:
\begin{eqnarray}
{\cal A}^{AATC} & \sim & \sum_{\rm non-cyclic}\int dx_{1}dx_{2}dx_{3}dzd\bar{z}\,
  \lan V_{A}^{(0)}{(x_{1})}
V_{A}^{(0)}{(x_{2})}V_T^{(-2)}{(x_{3})}
V_{RR}^{(0)}(z,\bar{z})\ran\labell{sfield}\eeqa
The internal  CP factor is $\Tr(II\sigma_1\sigma_1)=2$ for 123 ordering and  is $\Tr(I\sigma_1 I\sigma_1)=2$ for 132 ordering. The internal CP matrix of tachyon in $(-2)$-picture is the same as the CP matrix of this operator in $(0)$-picture which is $\sigma_1$. It is easier to calculate the S-matrix element in the following picture:
\begin{eqnarray}
{\cal A}^{AATC} & \sim & \sum_{\rm non-cyclic}\int dx_{1}dx_{2}dx_{3}dzd\bar{z}\,
  \lan V_{A}^{(0)}{(x_{1})}
V_{A}^{(0)}{(x_{2})}V_T^{(-1)}{(x_{3})}
V_{RR}^{(-1)}(z,\bar{z})\ran\labell{sstring}\eeqa
The  CP factor in this case is $\Tr(II\sigma_2\sigma_3\sigma_1)=2i$ for 123 ordering and is $\Tr(I\sigma_2 I\sigma_3\sigma_1)=2i$ for 132 ordering. According to the speculation in \cite{Garousi:2008ze},  the difference between the above two S-matrix elements is a factor of $i$ when  the internal CP factors are included. So to calculate the S-matrix element \reef{sfield} in which the vertexes carry the same internal CP matrices as the effective field theory, we first calculate the amplitude \reef{sstring} and then multiply the result by $i$. 

The vertex operators in \reef{sstring} are given as\footnote{In string theory side, our conventions  set $\alpha'=2$.}
\beqa
V_{T}^{(-1)}(y) &=& e^{-\phi(y)} e^{2ik\cd X(y)}\lam\otimes\sigma_2
\nonumber\\
V_{A}^{(0)}(x) &=& \xi_{i}\bigg(\partial X^i(x)+2ik\cd\psi\psi^i(x)\bigg)e^{2ik.X(x)}\lam\otimes I 
\nonumber\\
V_{RR}^{(-1)}(z,\bar{z})&=&(P_{-}\fsH_{(n)}M_p)^{\al\be}e^{-\phi(z)/2} S_{\al}(z)e^{ip\cd X(z)}e^{-\phi(\bar{z})/2} S_{\be}(\bar{z}) e^{ip\cd D \cd X(\bar{z})}\otimes\sigma_3\sigma_1\nonumber\eeqa
where $k$ is the momentum of open string that for tachyon satisfies the on-shell condition $k^2=1/4$, and $\lam$ is the external CP matrix  in the $U(N)$ group.  We refer the interested reader to ~\cite{Garousi:1996ad} for our conventions on the RR vertex operator. 
 
To calculate the correlators in \reef{sstring}, one must use the following standard propagators: 
\begin{eqnarray}
\lan X^{\mu}(z)X^{\nu}(w)\ran & = & -\eta^{\mu\nu}\log(z-w) , \non \\
\lan \psi^{\mu}(z)\psi^{\nu}(w) \ran & = & -\eta^{\mu\nu}(z-w)^{-1} \ ,\non \\
\lan\phi(z)\phi(w)\ran & = & -\log(z-w) \ ,
\labell{prop}\end{eqnarray}
Introducing $x_{4}\equiv\ z=x+iy$ and $x_{5}\equiv\bz=x-iy$,  the  amplitude  factorizes  to the following 
correlators for 123 ordering:\beqa {\cal A}^{AATC}&\sim& \int 
 dx_{1}dx_{2}dx_{3}dx_{4} dx_{5}\,
(P_{-}\fsH_{(n)}M_p)^{\al\be}\xi_{1i}\xi_{2j}x_{45}^{-1/4}x_{43}^{-1/2}x_{53}^{-1/2}(I_1+I_2+I_3+I_4)\nonumber\\&&
\times\Tr(\lam_1\lam_2\lam_3)\Tr(II\sigma_2\sigma_3\sigma_1)\labell{125}\eeqa
where $x_{ij}=x_i-x_j$ and 
\beqar I_1&=&{<:\partial X^i(x_1)e^{2ik_1.X(x_1)}:\partial X^j(x_2)e^{2ik_2.X(x_2)}
:e^{2ik_3.X(x_3)}:e^{ip.X(x_4)}:e^{ip.D.X(x_5)}:>}
 \  \non \\&&\times{<:S_{\al}(x_4):S_{\be}(x_5):>}\nonumber\\
I_2&=&{<:\partial X^i(x_1)e^{2ik_1.X(x_1)}:e^{2ik_2.X(x_2)}
:e^{2ik_3.X(x_3)}:e^{ip.X(x_4)}:e^{ip.D.X(x_5)}:>}
 \  \non \\&&\times{<:S_{\al}(x_4):S_{\be}(x_5):2ik_{2}\cd\psi\psi^j(x_2):>}\nonumber\\
 I_3&=&{<: e^{2ik_1.X(x_1)}:\partial X^j(x_2)e^{2ik_2.X(x_2)}
:e^{2ik_3.X(x_3)}:e^{ip.X(x_4)}:e^{ip.D.X(x_5)}:>}
 \  \non \\&&\times{<:S_{\al}(x_4):S_{\be}(x_5):2ik_{1}\cd\psi\psi^i(x_1):>}\nonumber\\
 I_4&=&{<: e^{2ik_1.X(x_1)}:e^{2ik_2.X(x_2)}
:e^{2ik_3.X(x_3)}:e^{ip.X(x_4)}:e^{ip.D.X(x_5)}:>}
 \  \non \\&&\times{<:S_{\al}(x_4):S_{\be}(x_5):2ik_{1}\cd\psi\psi^i(x_1):2ik_{2}\cd\psi\psi^j(x_2):>}\eeqar
One can perform easily the  correlators of $X$ using the corresponding propagator  in \reef{prop}. To find the correlator of $\psi$, we use the following Wick-like rule~\cite{Liu:2001qa} for the
 correlation function involving an arbitrary number of $\psi$'s 
and 
two $S$'s:
 \beqa
 \lan\psi^{\mu_1}
(y_1)...
\psi^{\mu_n}(y_n)S_{\al}(z)S_{\be}(\bz)\ran&\!\!\!\!=\!\!\!\!
&\frac{1}{2^{n/2}}
\frac{(z-\bz)^{n/2-5/4}}
{|y_1-z|...|y_n-z|}\left[(\Gamma^{\mu_n...\mu_1}
C^{-1})_{\al\be}\right.\nonumber\\&&+
\lan\lan\psi^{\mu_1}(y_1)\psi^{\mu_2}(y_2)\ran\ran(\Gamma^{\mu_n...\mu_3}
C^{-1})_{\al\be}
\pm perms\nonumber\\&&+\lan\lan\psi^{\mu_1}(y_1)\psi^{\mu_2}(y_2)\ran\ran
\lan\lan\psi^{\mu_3}(y_3)\psi^{\mu_4}(y_4)\ran\ran(\Gamma^{\mu_n...\mu_5}
C^{-1})_{\al\be}\nonumber\\&&\left.
\pm {\rm perms}+\cdots\right]\labell{wicklike}\eeqa where  dots mean  sum  over all possible contractions. In above equation, $\Gamma^{\mu_{n}...\mu_{1}}$ is the totally antisymmetric combination of the gamma matrices and  the Wick-like contraction is given by 
\beqa
\lan\lan\psi^{\mu}(y_{1})\psi^{\nu}(y_{2})\ran\ran &=&\eta^{\mu\nu}{\frac {(y_{1}-z)(y_{2}-\bz)+(y_{2}-z)(y_{1}-\bz)}{(y_{1}-y_{2})(z-\bz)}}\nonumber\\
&=&2\eta^{\mu\nu}{\frac {Re[(y_{1}-z)(y_{2}-\bz)]}{(y_{1}-y_{2})(z-\bz)}}
\eeqa 
where in the second line the fact that $y_1,\,y_2$ are real  has been used. One can use the above formula to find the correlation function of two $S$'s and an arbitrary number of currents. The only subtlety in using the  formula \reef{wicklike} for currents is that one must not consider the Wick-like contraction for the two $\psi$'s in one  current. Using this, one can easily  find the following standard results when there is no current and  when there is  one current:
\beqa  {<:S_{\al}(x_4):S_{\be}(x_5):>}&=&x_{45}^{-5/4}C^{-1}_{\al\be} \labell{exp22}\\
<:S_{\al}(x_4):S_{\be}(x_5):\psi^m\psi^i(x_1):>&=& -\frac{1}{2}x_{45}^{-1/4}
x_{14}^{-1}x_{15}^{-1}(\Gamma^{mi}C^{-1})_{\al\be}\nonumber\eeqa
In the second relation, we have not consider the Wick-like contraction for the two $\psi$'s since both of them belong to one current. When there are two currents, the formula \reef{wicklike} gives  the following result:
\beqa
I_4'&=&<:S_{\al}(x_4):S_{\be}(x_5):\psi^m\psi^i(x_1):\psi^l\psi^j(x_2):>\labell{63}\\
&=&\frac{1}{4}x_{45}^{3/4}(x_{14}x_{15}x_{24}x_{25})^{-1} \bigg\{(\Gamma^{jlim}C^{-1})_{\alpha\beta}+2\frac{Re[x_{14}x_{25}]}{x_{12}x_{45}}\bigg[\eta^{ml}(\Gamma^{ji}C^{-1})_{\alpha\beta}-\eta^{mj}(\Gamma^{li}C^{-1})_{\alpha\beta}\nonumber\\&&-\eta^{il}(\Gamma^{jm}C^{-1})_{\alpha\beta}+\eta^{ij}(\Gamma^{lm}C^{-1})_{\alpha\beta}\bigg]+4\bigg(\frac{Re[x_{14}x_{25}]}{x_{12}x_{45}}\bigg)^{2}
(-\eta^{ml}\eta^{ij}+\eta^{mj}\eta^{il})C^{-1}_{\alpha\beta}\bigg\}\nonumber\eeqa
Replacing the above spin correlators into \reef{125} and performing the correlators over $X$, one finds 
\beqa
{\cal A}^{AATC}&\!\!\!\!\sim\!\!\!\!\!&\int dx_{1}dx_{2} dx_{3}dx_{4}dx_{5}(P_{-}\fsH_{(n)}M_p)^{\al\be}I\xi_{1i}\xi_{2j}x_{45}^{-1/4}x_{43}^{-1/2}x_{53}^{-1/2}\nonumber\\&&\times
\bigg(x_{45}^{-5/4}C^{-1}_{\al\be}(-\eta^{ij}x_{12}^{-2}+a^i_1a^j_2)+a^i_1(a^j_3)_{\al\be}+a^j_2(a^i_4)_{\al\be}-4k_{1m}k_{2l}I_4'\bigg)\labell{amp3}\eeqa
where  $I_4'$ is given in \reef{63}, and
\beqa 
I&=&|x_{12}|^{4k_1.k_2}|x_{13}|^{4k_1.k_3}|x_{14}x_{15}|^{2k_1.p}|x_{23}|^{4k_2.k_3}|x_{24}x_{25}|^{2k_2.p}
|x_{34}x_{35}|^{2k_3.p}|x_{45}|^{p.D.p}\nonumber\\
a^i_1&=&-ik_2^{i}\bigg(\frac{x_{42}}{x_{41}x_{12}}+\frac{x_{52}}{x_{51}x_{12}}\bigg)
-ik_3^{i}\bigg(\frac{x_{43}}{x_{41}x_{13}}+\frac{x_{53}}{x_{51}x_{13}}\bigg)\nonumber\\
a^j_2&=&-ik_1^{j}\bigg(\frac{x_{14}}{x_{42}x_{12}}+\frac{x_{15}}{x_{52}x_{12}}\bigg)
-ik_3^{j}\bigg(\frac{x_{43}}{x_{42}x_{23}}+\frac{x_{53}}{x_{52}x_{23}}\bigg)\nonumber\\
(a^j_3)_{\al\be}&=&-ik_{2l}x_{45}^{-1/4}(\Gamma^{lj}C^{-1})_{\al\be}
(x_{24}x_{25})^{-1}\nonumber\\
(a^i_4)_{\al\be}&=&-ik_{1m}x_{45}^{-1/4}(\Gamma^{mi}C^{-1})_{\al\be}
(x_{14}x_{15})^{-1}\eeqa
 One can show that the integrand is invariant under
SL(2,R) transformation. Gauge fixing  this  symmetry by fixing the position of the open string vertex operators as  \beqar
 x_{1}&=&0 ,\qquad x_{2}=1,\qquad x_{3}\rightarrow \infty,
 \eeqar 
 One finds the following integral:
\beqa 
 \int d^2 \!z |1-z|^{a} |z|^{b} (z - \bar{z})^{c}
(z + \bar{z})^{d} 
 \eeqa
 where $d=0,1,2$ and $a,b,c$ are given in terms of the Mandelstam variables:
\beqar
s&=&-(k_1+k_3)^2,\qquad t=-(k_1+k_2)^2,\qquad u=-(k_2+k_3)^2
\qquad\eeqar  
The region of integration  is the upper half complex plane. For $d=0,1$ the result is given in  \cite{Fotopoulos:2001pt}, \ie
\beqa 
 \int d^2 \!z |1-z|^{a} |z|^{b} (z - \bar{z})^{c}
(z + \bar{z})^{d}&\!\!\!\!=\!\!\!&
(2i)^{c} 2^d \,  \pi \frac{ \Gamma( 1+ d +
\frac{b+c}{2})\Gamma( 1+ \frac{a+c}{2})\Gamma( -1-
\frac{a+b+c}{2})\Gamma( \frac{1+c}{2})}{
\Gamma(-\frac{a}{2})\Gamma(-\frac{b}{2})\Gamma(2+c+d+
\frac{a+b}{2})}\nonumber
\eeqa
Extending the result in \cite{Fotopoulos:2001pt} to $ d= 2$, one finds 
\beqa 
 \int d^2 \!z |1-z|^{a} |z|^{b} (z - \bar{z})^{c}
(z + \bar{z})^{d}&\!\!\!\!=\!\!\!&
(2i)^{c} 2^d \,  \pi \frac{J_1+J_2}{
\Gamma(-\frac{a}{2})\Gamma(-\frac{b}{2})\Gamma(d+2+c+
\frac{a+b}{2})}\nonumber\eeqa
where
\beqa 
J_1&=&\frac{1}{2}\Gamma( d+
\frac{b+c}{2})\Gamma( d+ \frac{a+c}{2})\Gamma( -d-
\frac{a+b+c}{2})\Gamma( \frac{1+c}{2})\nonumber\\
J_2&=&\Gamma( d+1+
\frac{b+c}{2})\Gamma( 1+ \frac{a+c}{2})\Gamma( -1-
\frac{a+b+c}{2})\Gamma( \frac{1+c}{2})\nonumber\eeqa
Using the above integrals, one can write the amplitude \reef{amp3} as 
\beqa {\cal A}^{AATC}&=&{\cal A}_{1}+{\cal A}_{2}+{\cal A}_{3}\labell{44}\eeqa 
where 
\beqa
{\cal A}_{1}&\!\!\!\sim\!\!\!&-2i\Tr(\lam_1\lam_2\lam_3)\xi_{1i}\xi_{2j}k_{1m}k_{2l}
\Tr(P_{-}\fsH_{(n)}M_p\Gamma^{jlim}
)(t+s+u+1/2)L_3 
\nonumber\\
{\cal A}_{2}&\sim&2\Tr(\lam_1\lam_2\lam_3)\bigg\{\bigg[k_{2l}\xi_{2j}\left(-2k_2.\xi_1 L_1+2k_3.\xi_1L_2\right)\Tr(P_{-}\fsH_{(n)}M_p \Gamma^{lj})\nonumber\\&&
-2k_1.\xi_2k_{2l}\xi_{1i}
\Tr(P_{-}\fsH_{(n)}M_p\Gamma^{il})L_1\bigg]-\bigg[1\leftrightarrow 2\bigg]\nonumber\labell{111}\\
&&-L_1\left(-t\xi_{1i}\xi_{2j}\Tr(P_{-}\fsH_{(n)}M_p\Gamma^{ji})
+2k_{2l}k_{1m}\xi_1.\xi_2\Tr(P_{-}\fsH_{(n)}M_p \Gamma^{lm}
)\right)
\bigg\}
\nonumber\\
{\cal A}_{3}&\sim&-2i\Tr(\lam_1\lam_2\lam_3)\Tr(P_{-}\fsH_{(n)}M_p)L_3\left[
-t(k_3.\xi_1)(k_3.\xi_2)+(k_3.\xi_2)(k_2.\xi_1)(s+\frac{1}{4})\right.
\nonumber\\&&+(k_3.\xi_1)(k_1.\xi_2)(u+\frac{1}{4})+\frac{1}{2}(\xi_1.\xi_2)(u+\frac{1}{4})(s+\frac{1}{4})\bigg]
\nonumber\eeqa
 The extra factor of $i$ in ${\cal A}_{2}$ is coming from the extra factor of $x_{45}=2iy$ in this amplitude. The functions  
 $L_1,L_2,L_3$ are the following :
\beqa 
L_1&=&(2)^{-2(t+s+u)-1}\pi{\frac{\Gamma(-u+\frac{3}{4})
\Gamma(-s+\frac{3}{4})\Gamma(-t)\Gamma(-t-s-u)}
{\Gamma(-u-t+\frac{3}{4})\Gamma(-t-s+\frac{3}{4})\Gamma(-s-u+\frac{1}{2})}}\nonumber\\
L_2&=&(2)^{-2(t+s+u)-1}\pi{\frac{\Gamma(-u+\frac{3}{4})
\Gamma(-s-\frac{1}{4})\Gamma(-t+1)\Gamma(-t-s-u)}
{\Gamma(-u-t+\frac{3}{4})\Gamma(-t-s+\frac{3}{4})\Gamma(-s-u+\frac{1}{2})}}\nonumber\\
L_3&=&(2)^{-2(t+s+u)}\pi{\frac{\Gamma(-u+\frac{1}{4})
\Gamma(-s+\frac{1}{4})\Gamma(-t+\frac{1}{2})\Gamma(-t-s-u-\frac{1}{2})}{\Gamma(-u-t+\frac{3}{4})
\Gamma(-t-s+\frac{3}{4})\Gamma(-s-u+\frac{1}{2})}}\nonumber\eeqa
 From the poles of the gamma functions, one realizes that the scattering amplitude has tachyon, massless and infinite number of massive poles. To compare the field theory which has tachyon and massless fields \eg the WZ action,  with the above amplitude, one must expand the amplitude such that the tachyon and massless poles of the field theory survive  and all other poles disappear in the form of contact terms. In the next section we will find such expansion. 

\section{Momentum expansion}

 Using the  momentum conservation along the world volume of brane,  $k_1^{i} + k_2^{i}+k_3^{i}+p^{i} =0$, one finds the Mandelstam variables satisfy 
\beqa
s+t+u=-p_ip^i-1/4
\labell{cons}\eeqa
 In general, it has been argued in \cite{Garousi:2007si} that the momentum expansion of a S-matrix element should be around  $(k_i+k_j)^2\rightarrow 0$ and/or $k_i\inn k_j\rightarrow 0$.  The case $(k_i+k_j)^2\rightarrow 0$ is when there is  massless pole in $(k_i+k_j)^2$-channel. One can easily observe that the amplitude \reef{sfield} must have massless pole only in $(k_1+k_2)^2$-channel, so the momentum expansion must be around 
 \beqa
k_3.k_1\rightarrow 0,\qquad k_3.k_2\rightarrow 0,\qquad (k_1+k_2)^2\rightarrow 0 \nonumber\eeqa
Using the on-shell relations $k_1^2=k_2^2=0$ and $k_3^2=1/4$, one can rewrite it in terms of the Mandelstam variables as
\beqa
s\rightarrow -1/4,\qquad u\rightarrow -1/4,\qquad  t\rightarrow 0 \labell{point}\eeqa
The constraint \reef{cons} then indicates that $p_ip^i\rightarrow 1/4$ which is possible only for  euclidean brane. 
This is consistent with the observation  made in  \cite{Billo:1999tv,Garousi:2008tn} that the on-shell condition implies that  the S-matrix element can be evaluated only  for non-BPS SD-branes \cite{Gutperle:2002ai}.  

Expansion of the functions $L_1,L_2,L_3$ around the above point is
\beqa
L_1&=&-\pi^{3/2}\bigg(\frac{1}{t}\sum_{n=-1}^{\infty}b_n(u+s+1/2)^{n+1}\nonumber\\&&+\sum_{p,n,m=0}^{\infty}e_{p,n,m}t^{p}((s+1/4)(u+1/4))^{n}(s+u+1/2)^m\bigg)\nonumber\\
L_2&=&-\pi^{3/2}\bigg(\frac{1}{(s+1/4)}\sum_{n=-1}^{\infty}b_n(u+t+1/4)^{n+1}\nonumber\\&&+\sum_{p, n,m=0}^{\infty}e_{p,n,m}(s+1/4)^{p}(t(u+1/4))^{n}(t+u+1/4)^m\bigg)\labell{high}\\
L_3&=&-{\pi^{5/2}}\left( \sum_{n=0}^{\infty}c_n(s+t+u+1/2)^n\right.\nonumber\\
&&\left.+\frac{\sum_{n,m=0}^{\infty}c_{n,m}[(s+1/4)^n(u+1/4)^m +(s+1/4)^m(u+1/4)^n]}{(t+s+u+1/2)}\right.\nonumber\\
&&\left.+\sum_{p,n,m=0}^{\infty}f_{p,n,m}(s+t+u+1/2)^p[(s+u+1/2)^n((s+1/4)(u+1/4))^{m}]\right)\nonumber\eeqa
where the coefficients $b_n$ are exactly the coefficients that appear in the momentum expansion of the S-matrix element of one RR, one gauge field  and one tachyon vertex operators \cite{Garousi:2008tn}.  Some of the coefficients $b_n,\,e_{p,n,m},\,c_n,\,c_{n,m}$ and $f_{p,n,m}$ are
\beqa 
&&b_{-1}=1,\,b_0=0,\,b_1=\frac{1}{6}\pi^2,\,b_2=2\z(3)\nonumber\\
&&e_{2,0,0}=e_{0,1,0}=2\z(3),e_{1,0,0}=\frac{1}{6}\pi^2,e_{1,0,2}=\frac{19}{60}\pi^4,e_{1,0,1}=e_{0,0,2}=6\z(3)\nonumber\\
&&e_{0,0,1}=\frac{1}{3}\pi^2,e_{3,0,0}=\frac{19}{360}\pi^4,e_{0,0,3}=e_{2,0,1}=\frac{19}{90}\pi^4,e_{1,1,0}=e_{0,1,1}=\frac{1}{30}\pi^4\labell{577}\\
&&c_0=0,c_1=\frac{\pi^2}{3},c_2=4\xi(3),
\,c_{1,1}=\frac{\pi^2}{3},\,c_{0,0}=1\nonumber\\
&&c_{1,0}=c_{0,1}=0
,c_{3,0}=c_{0,3}=0\, 
\,c_{2,0}=c_{0,2}=\frac{\pi^2}{3},c_{1,2}=c_{2,1}=-8\xi(3)\nonumber\\
&&f_{0,1,0}=-\frac{2\pi^2}{3},\,f_{0,2,0}=-f_{1,1,0}=12\xi(3),f_{0,0,1}=4\xi(3)\, \nonumber
\eeqa
$L_1$ has massless pole in $t$-channel, $L_2$ has  tachyonic pole in $s$-channel  and $L_3$ has  tachyonic pole in $(s+t+u)$-channel. These poles must  be reproduced in field theory by appropriate couplings. The string amplitude \reef{44} is non-zero for $p=n+3$, $p=n-1$ and for $p=n+1$. Let us study each case separately.
\subsection{$p=n+3$ case}
This  is the simplest case to consider. Only  ${\cal A}_1$ in \reef{44} is non-zero. The trace in ${\cal A}_1$ is:
\beqa
\Tr\bigg(\fsH_{(n)}M_p\Gamma^{jlim}\bigg)&=&\pm\frac{32}{n!}\eps^{jlimi_{0}\cdots i_{p-4}}H_{i_{0}\cdots i_{p-4}}\nonumber\eeqa
We are going to compare string theory S-matrix elements with field theory S-matrix elements including their coefficients, however, we are not interested in fixing the overall sign of the amplitudes. Hence, in above and in the rest of equations in this paper, we have payed   no attention to the overall sign of  equations. The string amplitude for electric RR field then becomes  
\beqa
{\cal A}^{AATC}=\mp\frac{32i}{(p-3)!}(\mu'_p\beta'\pi^{1/2})\Tr(\lam_1\lam_2\lam_3)\xi_{1i}\xi_{2j}k_{1m}k_{2l}
\eps^{jlimi_{0}\cdots i_{p-4}}H_{i_{0}\cdots i_{p-4}}(t+s+u+1/2)L_3\nonumber\eeqa 
where we have also normalized the amplitude by  $(\mu'_p\beta'\pi^{1/2})$.  Apart from the group factor the above amplitude is antisymmetric under interchanging the gauge fields . So the whole amplitude  is zero for abelian gauge group. The   amplitude also satisfies the Ward identity, \ie the amplitude vanishes under replacing  each of  $\xi^i\rightarrow k^i$. Since  $(t+s+u+1/2)L_3$ has no tachyon/massless pole, the amplitude  has only contact terms. The leading contact term is reproduced by the following  coupling:  
\beqa
\beta'\mu_p'(2\pi\alpha')^{3}\Tr (C_{p-4}\wedge F\wedge F\wedge DT)\labell{hderv}
\eeqa
and the non-leading order  terms should be corresponding   to the higher derivative extension of the above coupling. This coupling is exactly given by the WZ terms \reef{WZ'} after expanding the exponential and using the multiplication rule of the supermatrices \cite{Garousi:2008tn}.
\subsection{$p=n-1$ case}

The next simple case to consider is  $p=n-1$. Only ${\cal A}_3$ in \reef{44} is non-zero  for this case. The  trace in this amplitude is:
\beqa
\Tr\bigg(\fsH_{(n)}M_p
\bigg)&=&\pm\frac{32}{n!}\eps^{i_{0}\cdots i_{p}}H_{i_{0}\cdots i_{p}}
\nonumber\eeqa
Substituting  this trace in  ${\cal A}_3$,  one finds 
\beqa
{\cal A}^{AATC}&=&\mp\frac{32i}{(p+1)!}(\beta'\mu'_p\pi^{1/2})\Tr(\lambda_1\lambda_2\lambda_3)\eps^{i_{0}\cdots i_{p}}H_{i_{0}\cdots i_{p}}L_3\bigg\{
-t(k_3.\xi_1)(k_3.\xi_2)
\labell{Aaatc}\\&&+(k_3.\xi_2)(k_2.\xi_1)(s+\frac{1}{4})+(k_3.\xi_1)(k_1.\xi_2)(u+\frac{1}{4})+\frac{1}{2}(\xi_1.\xi_2)(u+\frac{1}{4})(s+\frac{1}{4})
\bigg\}\nonumber\eeqa
where we have also normalized the amplitude by $(\beta'\mu'_p\pi^{1/2})$. The amplitude satisfies the Ward identity and it is symmetric under interchanging the gauge fields. So the amplitude is non-zero even for abelian case. 

All terms in \reef{Aaatc} have tachyon pole in the $(s+t+u)$-channel and infinite contact terms. We consider only the tachyon pole and show that they can be reproduced by WZ coupling $C_{p}\wedge DT$ and the higher derivative two-gauge-two-tachyon couplings that have been found in \cite{Garousi:2008xp}. 
 To this end,  consider  the amplitude for decaying   one R-R field  to two gauge fields and one tachyon in the world-volume theory  of the non-BPS branes which is given by  the following  Feynman amplitude :  
\beqa
{\cal A}&=&V^{\alpha}(C_{p},T)G^{\alpha\beta}(T)V^{\beta}(T,T_3,A_1,A_2)\labell{amp544}\eeqa
where the tachyon propagator and the vertex $V^{\alpha}(C_{p},T)$ are given as 
\beqa
G^{\alpha\beta}(T) &=&\frac{i\delta^{\alpha\beta}}{(2\pi\alpha') T_p
(-k^2-m^2)}\nonumber\\
V^{\alpha}(C_{p},T)&=&2i\mu'_p\beta'(2\pi\alpha')\frac{1}{(p+1)!}\epsilon^{i_0\cdots i_{p}}H_{i_0\cdots i_{p}}\Tr(\Lambda^{\alpha})
\labell{Fey}
\eeqa
In above vertex, $\Tr(\Lambda^{\alpha})$ is non-zero only for abelian matrix $\Lambda^{\alpha}$. The vertex $ V^{\beta}(T,T_3,A_1,A_2)$  can be  derived from the higher derivative  of two-gauge-two-tachyon couplings \cite{Garousi:2008xp} (equation (29) of \cite{Garousi:2008xp}). They are the higher derivative extension of two-gauge-two-tachyon couplings of the tachyon action \reef{nonab}. Using the fact that the off-shell tachyon is abelian, one finds  the vertex $V^{\beta}(T,T_3,A_1,A_2)$ to be 
\beqa
&&2iT_p(\pi\alpha')(\alpha')^{2+n+m}
(a_{n,m}+b_{n,m})\Tr(\lambda_1\lambda_2\lambda_3\Lambda^{\beta})\bigg[-t(k_3.\xi_1)(k_3.\xi_2)+(k_3.\xi_2)(k_2.\xi_1)
(s+\frac{1}{4})\nonumber\\&&+(k_3.\xi_1)(k_1.\xi_2)(u+\frac{1}{4})+\frac{1}{2}(\xi_1.\xi_2)(u+\frac{1}{4})(s+\frac{1}{4})\bigg]\bigg(\frac{}{}(k_3\inn k_1)^n(k_3\inn k_2)^m+(k_3\inn k_1)^n(k_1\inn k)^m \nonumber\\&&
+(k\inn k_2)^m(k\inn k_1)^n+(k_1\inn k)^n(k_3\inn k_1)^m+(k_3\inn k_2)^m(k_2\inn k)^n+(k\inn k_2)^n(k_1\inn k)^m
+(k_3\inn k_2)^n\nonumber\\&&\times(k_1\inn k_3)^m+(k_3\inn k_2)^n(k_2\inn k)^m\bigg)\labell{veraatt}\eeqa
where  $k$ is the momentum of the off-shell tachyon. There are similar terms which have coefficient $\Tr(\lambda_2\lambda_1\lambda_3\Lambda^{\beta})$. Some of the coefficients $a_{n,m}$ and $b_{n,m}$ are \cite{Garousi:2008xp}
\beqa
&&a_{0,0}=-\frac{\pi^2}{6},\,b_{0,0}=-\frac{\pi^2}{12}\\
&&a_{1,0}=2\z(3),\,a_{0,1}=0,\,b_{0,1}=b_{1,0}=-\z(3)\nonumber\\
&&a_{1,1}=a_{0,2}=-7\pi^4/90,\,a_{2,0}=-4\pi^4/90,\,b_{1,1}=-\pi^4/180,\,b_{0,2}=b_{2,0}=-\pi^4/45\nonumber\\
&&a_{1,2}=a_{2,1}=8\z(5)+4\pi^2\z(3)/3,\,a_{0,3}=0,\,a_{3,0}=8\z(5),\nonumber\\
&&\qquad\qquad\qquad\qquad\qquad b_{0,3}=-4\z(5),\,b_{1,2}=-8\z(5)+2\pi^2\z(3)/3\nonumber\eeqa
and $b_{n,m}$ is symmetric.

Now one can write  $k_1\inn k=k_2.k_3-(k^2+m^2)/2$ and $k_2\inn k=k_1.k_3-(k^2+m^2)/2$. 
The terms $k^2+m^2$ in the  vertex \reef{veraatt} will be canceled with the $k^2+m^2$ in the denominator of the tachyon propagator resulting a bunch of contact terms of one RR, two gauge fields and one tachyon in which we are not interested. Ignoring them,   one finds the following tachyon pole :
\beqa
&&-32\pi\alpha'^{2}\beta'\mu_p'\frac{ \eps^{i_{0}\cdots i_{p}}H_{i_{0}\cdots i_{p}}}{(p+1)!(s'+t+u')}\Tr(\lam_1\lam_2\lam_3)
\sum_{n,m=0}^{\infty}\bigg((a_{n,m}+b_{n,m})[s'^{m}u'^{n}+s'^{n}u'^{m}]\nonumber\\&&\times
\bigg[-t(k_3.\xi_2)(k_3.\xi_1)+(k_2.\xi_1)(k_3.\xi_2)s'+(k_1.\xi_2)(k_3.\xi_1)u'+(\xi_1.\xi_2)\frac{1}{2}u's'
\bigg]\bigg)\label{amphigh}\eeqa 
where  $u'=u+1/4=-\alpha'k_2\inn k_3$ and $s'=s+1/4=-\alpha'k_1\inn k_3$.
The above amplitude should be compared with the tachyon pole in  \reef{Aaatc}. 
Let us compare them for some values of $n,m$. For $n=m=0$, the amplitude \reef{amphigh} has the following numerical factor:
\beqa
-8(a_{0,0}+b_{0,0})&=&-8(\frac{-\pi^2}{6}+\frac{-\pi^2}{12})=2\pi^2\nonumber\eeqa
Similar term in \reef{Aaatc} has the numerical factor  $(2\pi^2c_{0,0})$ which is equal to the above number.  At the order of $\alpha'$, the amplitude \reef{amphigh} has the following numerical factor:
\beqa
-4(a_{1,0}+a_{0,1}+b_{1,0}+b_{0,1})(s'+u')&=&0\nonumber\eeqa
 Similar term in \reef{Aaatc} is proportional to   $\pi^2c_{1,0}(s+u+1/2)$ which is zero.  At the  order of $(\alpha')^2$, the amplitude \reef{amphigh} has the following factor :
\beqa
&&-8(a_{1,1}+b_{1,1})(s')(u')-4(a_{0,2}+a_{2,0}+b_{0,2}+b_{2,0})[(s')^2+(u')^2]\nonumber\\
&&=\frac{\pi^4}{3}(2s'u')+\frac{2\pi^4}{3}(s'^2+u'^2)
\nonumber\eeqa
Similar term in \reef{Aaatc} has numerical factor  $\pi^2c_{1,1}(2s'u')+\pi^2(c_{2,0}+c_{0,2})(s'^2+u'^2)$ which is equal to the above factor using the coefficients \reef{577}.  
At the order of $\alpha'^3$, this amplitude has the following factor :
\beqa
&&-4(a_{3,0}+a_{0,3}+b_{0,3}+b_{3,0})[(s')^3+(u')^3]-4(a_{1,2}+a_{2,1}+b_{1,2}+b_{2,1})[(s')(u')(s'+u')]\nonumber\\
&&=-16\pi^2\xi(3)s'u'(s'+u')
\nonumber\eeqa
which is equal to corresponding term in \reef{Aaatc}, \ie $\pi^2(c_{0,3}+c_{3,0})[(s')^3+(u')^3]+\pi^2(c_{2,1}+c_{1,2})s'u'(s'+u')$.  
Similar comparison can be done for all order of $\alpha'$. Hence, the field theory amplitude \reef{amphigh} reproduces  exactly the infinite tower of the tachyon pole of string theory amplitude \reef{Aaatc}. This indicates that the momentum expansion of the amplitude $CAAT$ is consistent with the momentum expansion of the amplitude $TTAA$ found in \cite{Garousi:2008xp}. 


\subsection{$p=n+1$ case}
We finally  consider the case  $p=n+1$. Only ${\cal A}_2$ in \reef{44} is non-zero  for this case. The trace in this amplitude is: 
\beqa
\Tr\bigg(\fsH_{(n)}M_p
\Gamma^{ij}\bigg)&=&\pm\frac{32}{n!}\eps^{i_{0}\cdots i_{p-2} ij}H_{i_{0}\cdots i_{p-2}}
\nonumber\eeqa
Substituting this trace in ${\cal A}_2$, one finds 
\beqa
{\cal A}^{AATC}&=&\mp\frac{32}{(p-1)!}(\mu'_p\beta'\pi^{1/2})\Tr(\lam_1\lam_2\lam_3)H_{i_{0}\cdots i_{p-2}}\eps^{i_{0}\cdots i_{p}}\bigg\{ 
\bigg(2k_2.\xi_1 k_{2i_{p-1}}\xi_{2i_{p}}\nonumber\\&&-2k_1.\xi_2k_{1i_{p-1}}\xi_{1i_{p}}+2k_1.\xi_2\xi_{1i_{p-1}}k_{2i_{p}}+2k_2.\xi_1
\xi_{2i_{p}}k_{1i_{p-1}}-t\xi_{1i_{p}}\xi_{2i_{p-1}}\nonumber\\&&+2\xi_1.\xi_2 k_{1i_{p}}k_{2i_{p-1}}\bigg)L_1+\bigg(-2k_3.\xi_1k_{2i_{p-1}}\xi_{2i_{p}}L_2-1\leftrightarrow 2\bigg)
\bigg\}\labell{line9}\eeqa
where again we have normalized the amplitude by $(\mu'_p\beta'\pi^{1/2})$. Apart
 from the group factor the amplitude is antisymmetric under interchanging the gauge fields, so the whole amplitude is zero for abelian gauge group. The amplitude  satisfies the Ward identity. The first six terms have  contact terms as well as  massless pole in t-channel and the last two terms in \reef{line9} have contact terms as well as  tachyon poles in s-channel and u-channel. We are going to analyze all order of the tachyon/massless poles and the leading order and next to the leading order contact terms in this section.  Let us study  each case separately.

\subsubsection{Tachyon pole }
We first consider the   tachyon pole. Replacing \reef{high} in above amplitude, one finds the following tachyon poles:

\beqa
{\cal A}^{AATC}&=&\mp\frac{32}{(p-1)!}(2\pi^{3/2})(\mu'_p\beta'\pi^{1/2})\Tr(\lam_1\lam_2\lam_3) H_{i_{0}\cdots i_{p-2}} 
\eps^{i_{0}\cdots i_{p}}\labell{Fey36}\\&&\times\sum_{n=-1}^{\infty}b_n\bigg(\frac{(u+t+1/4)^{n+1}}{s+1/4}(k_3.\xi_1)k_{2i_{p-1}}\xi_{2i_{p}}-\frac{(s+t+1/4)^{n+1}}{u+1/4}(k_3.\xi_2)k_{1i_{p-1}}\xi_{1i_{p}}\bigg)\nonumber\eeqa
 and some contact terms that we consider them  in section 3.3.3. Since \reef{Fey36} is antisymmetric under interchanging  $1 \leftrightarrow 2$,  we consider only the first term. This term should be reproduced in field theory by the following Feynman amplitude:
\beqa
{\cal A}&=&V^{\alpha}(C_{p-2},A_2,T)G^{\alpha\beta}(T)V^{\beta}(T,T_3,A_1)\labell{amp42}\eeqa
where the vertices can be found from the standard nonabelian kinetic term of the tachyon and from the higher derivative extension of the WZ coupling $C_{p-2}\wedge F\wedge DT$ found in \cite{Garousi:2008tn} (equation (16) of \cite{Garousi:2008tn}), \ie 
\beqa
V^{\beta}(T,T_3,A_1)&\!\!\!\!=\!\!\!\!&iT_p(2\pi\alpha')(k_3-k).\xi_1\Tr(\lam_3\lam_1\Lambda^\beta)\nonumber\\
V^{\alpha}(C_{p-2},A_2,T)&\!\!\!\!=\!\!\!\!&2\mu'_p\beta'\frac{(2\pi\alpha')^{2}}{(p-1)!}\epsilon^{i_0\cdots i_{p}}H_{i_0\cdots i_{p-2}}k_{2i_{p-1}}\xi_{2i_{p}}\sum_{n=-1}^{\infty}b_n(\alpha'k_2\cdot k)^{n+1}\Tr(\lam_2\Lambda^{\alpha})
\nonumber\eeqa
where $k$ is the momentum of the off-shell tachyon. Note that  the vertex $V^{\beta}(T,T_3,A_1)$ has no higher derivative correction as it arises from the kinetic term of the tachyon. The amplitude \reef{amp42} then becomes
\beqa
{\cal A}&=&4\mu'_p\beta'(2\pi\alpha')^{2}\frac{1}{(p-1)!(s+\frac{1}{4})}\Tr(\lam_1\lam_2\lam_3)\eps^{i_{0}\cdots i_{p}}H_{i_{0}\cdots i_{p-2}} k_{2i_{p-1}}\xi_{2i_{p}}(k_3.\xi_1)\nonumber\\&&\times\sum_{n=-1}^{\infty}b_n\bigg(\frac{\alpha'}{2}\bigg)^{n+1}(t+u+1/4)^{n+1}\nonumber\eeqa
which is exactly the tachyon pole of the string theory amplitude \reef{Fey36}.
 
 \subsubsection{Massless pole }
We now consider the   massless pole.  Replacing the expansion of $L_1$   into \reef{line9}, one finds the following massless pole in $t$-channel :
\beqa
{\cal A}^{AATC}&=&\pm\frac{32\mu'_p\beta'\pi^{2}}{t(p-1)!}\Tr(\lam_1\lam_2\lam_3)H_{i_{0}\cdots i_{p-2}}
\eps^{i_{0}\cdots i_{p}}\sum_{n=-1}^{\infty}b_n(u+s+1/2)^{n+1}\bigg[2k_2.\xi_1 k_{2i_{p-1}}\xi_{2i_{p}}\nonumber\\&&-2k_1.\xi_2k_{1i_{p-1}}\xi_{1i_{p}}+2k_1.\xi_2\xi_{1i_{p-1}}k_{2i_{p}}+2k_2.\xi_1\xi_{2i_{p}}k_{1i_{p-1}}+2\xi_1.\xi_2k_{1i_{p}}k_{2i_{p-1}}
\bigg]
\labell{masspole}\eeqa
and some contact terms that we consider them in section 3.3.3.  In field theory, the massless pole is given by the following  Feynman amplitude :
\beqa
{\cal A}&=&V_\alpha^{i}(C_{p-2},T_3,A)G_{\alpha\beta}^{ij}(A)V_\beta^{j}(A,A_1,A_2)\labell{amp54}\eeqa
The vertices and propagator are 
\beqa
V_\alpha^{i}(C_{p-2},T_3,A)&=&2\mu'_p\beta'(2\pi\alpha')^{2}\frac{1}{(p-1)!} \eps^{i_{0}\cdots i_{p-1}i}H_{i_{0}\cdots i_{p-2}}k_{i_{p-1}}\sum_{n=-1}^{\infty}b_n(\alpha'k_3\cdot k)^{n+1}\Tr(\lam_3\Lambda^{\alpha})\nonumber\\
V_\beta^{j}(A,A_1,A_2)&=&-iT_p(2\pi\alpha')^{2}\Tr(\lambda_{1}\lambda_{2}\Lambda_{\beta})[\xi_{1}^{j}(k_1-k).\xi_{2}
+\xi_{2}^{j}(k-k_2).\xi_{1}+\xi_{1}.\xi_{2}(k_2-k_1)^{j}]\nonumber\\
G_{\alpha\beta}^{ij}(A)&=&\frac{i\delta_{\alpha\beta}\delta^{ij}}{(2\pi\alpha')^{2}T_p(t)}\nonumber\eeqa
where $k$ is  momentum of the  off-shell gauge field. Here again the vertex $V_\alpha^{i}(C_{p-2},T_3,A)$ has been found from the higher derivative extension of the WZ coupling $C_{p-2}\wedge F\wedge DT$ that has been found in \cite{Garousi:2008tn}. Note again that  the vertex $V_\beta^{j}(A,A_1,A_2)$ has no higher derivative correction as it arises from the kinetic term of the gauge field. Replacing them in the amplitude \reef{amp54}, one finds
\beqa
{\cal A}&=&(2\pi\alpha')^{2}\frac{2\mu'_p\beta'}{(p-1)!t}\eps^{i_{0}\cdots i_{p-1}i}H_{i_{0}\cdots i_{p-2}}
\Tr(\lambda_{1}\lambda_{2}\lambda_{3})\sum_{n=-1}^{\infty}b_n\bigg(\frac{\alpha'}{2}\bigg)^{n+1}(s+u+1/2)^{n+1}
\nonumber\\&&\times\bigg(2(k_2.\xi_1)k_{1i_{p-1}}\xi_{2i}-2(k_1.\xi_2)k_{1i_{p-1}}\xi_{1i}-2(k_1.\xi_2)k_{2i_{p-1}}\xi_{1i}+2(k_2.\xi_1)\xi_{2i}k_{2i_{p-1}}\nonumber\\&&-2(\xi_1.\xi_2)k_{1i_{p-1}}k_{2i}\bigg)\labell{ver22}\eeqa
where we have used   $\sum_{\alpha}\lambda_{ij}^{\alpha}\lambda_{kl}^{\alpha}=\delta_{ik}\delta_{jl}$.
This is exactly the massless pole of the string theory amplitude \reef{masspole}. This indicates that the momentum expansion of the S-matrix element $CAAT$ in this paper is consistent with the momentum expansion of the S-matrix element $CAT$ found in \cite{Garousi:2008tn}.

\subsubsection{Contact terms} 
Replacing \reef{high}  into \reef{line9}, one finds the following contact terms at leading order and next to the leading order:
\beqa
{\cal A}^{AATC}&=&\mp\frac{32}{(p-1)!}(\mu'_p\beta'\pi^{2})\Tr(\lam_1\lam_2\lam_3)H_{i_{0}\cdots i_{p-2}}\eps^{i_{0}\cdots i_{p}} \bigg\{\xi_{1i_{p}}\xi_{2i_{p-1}}-\frac{\pi^2}{6}
\bigg(2k_2.\xi_1 k_{2i_{p-1}}\xi_{2i_{p}} \nonumber\\&&-2k_1.\xi_2 k_{1i_{p-1}}\xi_{1i_{p}}+2k_1.\xi_2\xi_{1i_{p-1}}k_{2i_{p}}+2k_2.\xi_1\xi_{2i_{p}}k_{1i_{p-1}}-t\xi_{1i_{p}}\xi_{2i_{p-1}}\nonumber\\&&+2\xi_1.\xi_2k_{1i_{p}} k_{2i_{p-1}}\bigg)\bigg[t+2(s+u+1/2)\bigg]+\frac{\pi^2}{6}
\xi_{1i_{p}}\xi_{2i_{p-1}}(s+u+1/2)^{2}\nonumber
\\&&
+\bigg(\frac{\pi^2}{3}k_3.\xi_1k_{2i_{p-1}}\xi_{2i_{p}}\bigg[2(t+u+1/4)+s+1/4\bigg]-[1\leftrightarrow 2]\bigg)\bigg\}
\labell{6}\eeqa
The first term is reproduced by $CAAT$ coupling of the following  gauge invariant coupling:
\beqa
2\beta'\mu_p'(2\pi\alpha')^2\Tr(C_{p-2}\wedge F\wedge DT)
\eeqa
which  is exactly given by the WZ terms \reef{WZ'} after expanding the exponential and using the multiplication rule of the supermatrices \cite{Garousi:2008tn}. The other terms in \reef{6} should be related to the higher derivative extension of the above coupling. However, there are many other higher derivative gauge invariant couplings which have contribution to the contact terms of the S-matrix element of $CAAT$. Comparing them with the string theory contact terms \reef{6}, one can not fix their coefficients uniquely.  One particular set of higher derivative gauge invariant   couplings that  reproduce the contact terms in \reef{6} are the following:
\beqa
-\frac{1}{12}\beta'\mu_p'(2\pi\alpha')^4\!\!\!\!\!\!\!\!&\bigg[-iD^{\beta}F_{a\alpha }D^{\alpha}F_{b\beta}D_cT+\frac{3i}{2}F_{ac}D_{\alpha}F_{\beta b}D^{\alpha}D^{\beta}T-\frac{3i}{2}D_{\alpha}F_{\beta b }F_{ac}D^{\alpha}D^{\beta}T\nonumber\\&
-\frac{1}{2}D_aD^{\alpha}D_cF_{b\alpha}D_{\beta}D^{\beta}T +F_{a\alpha }D^{\beta}D^{\alpha}D_{\beta}D_bD_cT
-\frac{1}{2}D_aD^{\alpha}D_{\beta}D^{\beta}F_{b\alpha}D_cT&\nonumber\\&+D_bD_cF_{a\alpha}D^{\beta} D^{\alpha}D_{\beta}T
+4D^{\alpha}D_a D_cF_{\beta b}D_{\alpha}D^{\beta}T-\frac{1}{2}D_aF_{\alpha \beta}D_bD^{\alpha}D^{\beta}D_cT&\nonumber\\
&-D_aD^{\beta}D_{\beta}D_cF_{b\alpha}D^{\alpha}T+2D_bD^{\alpha}D^{\beta}F_{a\alpha}D_{\beta}D_cT+D^{\alpha}D_{\alpha}D_cF_{\beta b }D^{\beta}D_aT&\nonumber\\
&+D_aD^{\beta}D_{\beta}F_{ b\alpha }D^{\alpha}D_cT+\frac{1}{2}D^{\beta}D^{\alpha}D_{\beta}D_cF_{a\alpha}D_b T&\nonumber\\
&-\frac{1}{2}D^{\alpha}D^{\beta}F_{ab} D_{\alpha}D_{\beta}D_cT\bigg]
\frac{1}{(p-2)!}C_{i_{0}\cdots i_{p-3}}\eps^{i_{0}\cdots i_{p-3}abc}&\labell{699}
\eeqa
where $D_aT=\prt_aT-i[A_a,T]$.
Among the couplings in \reef{699}, only  the last coupling  has non-zero on-shell $CTA$ coupling. This coupling has been found in \cite{Garousi:2008tn} from the S-matrix element of one RR, one gauge and one tachyon vertex operators. This coupling has been also used in the previous section to verify that the tachyon/massless poles in \reef{line9} are reproduced by the higher derivative couplings in field theory. All coupling in \reef{699} are at $(\alpha')^4$ order. The next order terms should be at $(\alpha')^5$ order, and so on. 

As we have mentioned in the Introduction section, the WZ couplings can also be found using the BSFT. In that framework, it has been argued in \cite{Kraus:2000nj} when the RR field is constant, there is no higher derivative correction to the WZ couplings. So one may expect that the above higher derivative WZ couplings should be zero for constant RR field. However, as we have mentioned before, the above couplings are valid when $p_ip^i\rightarrow 1/4$. So they can not be compared with the $p_ip^i=0$ result of  the BSFT.


\section*{Acknowledgment}

 E.H would like to thank A.Ghodsi for useful comments.


\end{document}